\documentclass[12pt,onecolumn,english]{IEEEtran}
\IEEEoverridecommandlockouts
\usepackage[T1]{fontenc}
\usepackage[latin9]{inputenc}
\usepackage{float}
\usepackage{amsmath}
\usepackage{amsthm}
\usepackage{amssymb}
\usepackage{graphicx}
\usepackage[unicode=true,
 bookmarks=false,
 breaklinks=false,pdfborder={0 0 1},backref=section,colorlinks=false]
 {hyperref}
\hypersetup{ colorlinks,linkcolor=blue,filecolor=blue,citecolor=blue,urlcolor=Darkblue}

\makeatletter

\floatstyle{ruled}
\newfloat{algorithm}{tbp}{loa}
\providecommand{\algorithmname}{Algorithm}
\floatname{algorithm}{\protect\algorithmname}

\theoremstyle{plain}
\newtheorem{thm}{\protect\theoremname}
\theoremstyle{definition}
\newtheorem{example}[thm]{\protect\examplename}

\usepackage{babel}
\usepackage{url}

\usepackage{tikz,hyperref}
\usepackage{orcidlink}
\definecolor{Darkblue}{rgb}{0.0, 0.0, 0.55} % RGB values for Dark Blue

\ifdefined\showcaptionsetup
 \PassOptionsToPackage{caption=false}{subfig}
\fi
\usepackage{subfig}
\makeatother

\usepackage{babel}
\providecommand{\examplename}{Example}
\providecommand{\theoremname}{Theorem}

\begin{document}
\title{TFDWT: Fast Discrete Wavelet Transform TensorFlow Layers}
\author{\IEEEauthorblockN{Kishore K. Tarafdar\orcidlink{0009-0001-4548-7639} and Vikram M.
Gadre\orcidlink{0000-0001-8439-5625}} \\
 \IEEEauthorblockA{Department of Electrical Engineering, Indian Institute of Technology
Bombay }}
\maketitle

% \tableofcontents{}
% \cleardoublepage{}

\begin{abstract}
TFDWT is an open-source Python library that allows the construction of TensorFlow Layers for Fast Discrete Wavelet Transform (DWT) and Inverse Discrete Wavelet Transform (IDWT) in end-to-end backpropagation learning networks. By definition, a multiresolution signal representation using a multi-rate discrete wavelet system creates enriched joint natural-frequency representations. These layers facilitate the construction of multilevel DWT filter banks and Wavelet Packet Transform (WPT) filter banks for a spatial-frequency representation of the inputs and features in shallow or deep networks. The discrete wavelet system partitions the frequency plane into subbands using orthogonal dilated and translated lowpass scaling and highpass wavelet function. The realization of a fast discrete wavelet system is a two-band perfect reconstruction multi-rate filter bank with FIR filters corresponding to the impulse responses of the scaling and wavelet function with downsampling and upsampling operations. A filter bank for a higher dimensional input is a seamless extension by successive separable circular convolutions across each independent axis. The command `pip install TFDWT' installs the latest version of the package.
\end{abstract}

\section{Introduction}

The multiresolution tiling of the frequency plane with localized natural-frequency domain features is often used to represent data to build economical and explainable models. A Discrete Wavelet Transform (DWT) is a widely used multiresolution transform for representing discrete data using an orthogonal or biorthogonal basis. The
Wavelet toolbox \cite{misiti1996wavelet} by MathWorks is a proprietary
software that has served the requirements for $\text{D}$-dimensional
wavelet transforms in the MATLAB environment for a few decades. Several
open-source packages are now available for 1D and 2D DWT in Python.
Pywavelets \cite{lee2019pywavelets} is a $\text{D}$-dimensional
wavelet transform library in Python that works with Numpy \cite{harris2020array}
arrays. However, it is challenging to directly use Pywavelets with
the symbolic tensors in TensorFlow \cite{tensorflow2015-whitepaper}
layers and CUDA \cite{fatica2008cuda}. WaveTF \cite{versaci2021wavetf}
is a solution for constructing 1D and 2D DWT layers in TensorFlow
but is limited to only Haar and Db2 wavelets. The package tensorflow-wavelets
\cite{tensorflow-wavelets-1.1.2} supports 1D and 2D transforms withmultiple wavelets. In Pytorch \cite{imambi2021pytorch}, the pytorch-wavelets \cite{pytorch-wavelets-1.3.0}package
allows the construction of 1D and 2D DWT layers. However, limited libraries are available for 3D and higher dimensional transforms for parallel Graphics Processing Unit (GPU) computations for a wide range of wavelet families.

% , but it has a minor bug in perfect reconstruction due to the padding and boundary effects in processing the finite-length inputs. 

\smallskip{}

For a $\text{D}$-dimensional wavelet $\boldsymbol{\psi}\in L^{2}(\mathbb{R})^{\text{{D}}}$,
a discrete wavelet system defined by $\big\{\boldsymbol{\psi}_{m,\boldsymbol{p}}:m\in\mathbb{Z},\boldsymbol{p}\in\mathbb{Z}^{\text{{D}}},\text{D}\in\mathbb{N}\big\}$
forms an orthogonal basis, where $\boldsymbol{\psi}_{m,\boldsymbol{p}}(\boldsymbol{x}):=2^{m}\psi(2^{m}\boldsymbol{x}-\boldsymbol{p})$.
Then, by definition the DWT of $\boldsymbol{x}\in\mathbb{Z}^{\text{{D}}}$
is $\boldsymbol{x}\mapsto(\langle\boldsymbol{x},\psi_{m,\boldsymbol{p}}\rangle)_{m,\boldsymbol{p}}$,
where $m$ is the dilation parameter and $\boldsymbol{p}$ is the
shift or translation parameter. The TFDWT Python package is a simple standalone DWT and IDWT library with minimal dependencies for computation with symbolic tensors and CUDA in modern deep learning frameworks. This paper concisely states the mathematics of realizing fast $\text{D}$-dimensional DWT and IDWT layers with filter bank structures with multichannel batched tensors. The current TFDWT release is a TensorFlow 2 realization with support up to fast DWT3D and IDWT3D with various orthogonal and biorthogonal wavelet families having impulse responses of diverse lengths. This paper also serves as a manual for seamlessly upgrading to higher dimensional separable transforms of the independent axes and porting the codes into other tensor processing frameworks with minimal software engineering.

 % The boundary effects are taken care of with cyclic convolutions instead of padding. 

\section{Discrete wavelet system for sequences}

\begin{figure}[h]
\begin{centering}
\subfloat[bior3.1 analysis (left), synthesis (right)]{\begin{centering}
\includegraphics[scale=0.36]{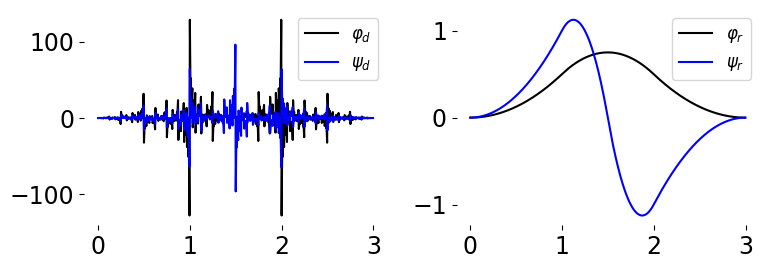}
\par\end{centering}
} \subfloat[bior3.1]{\begin{centering}
`\includegraphics[scale=0.42]{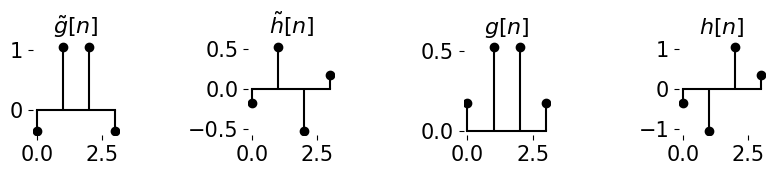}
\par\end{centering}
}
\par\end{centering}
\begin{centering}
\subfloat[rbio3.1 analysis (left), synthesis (right)]{\begin{centering}
\includegraphics[scale=0.36]{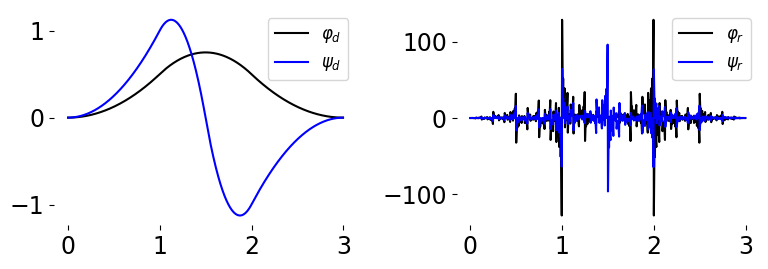}
\par\end{centering}
} \subfloat[rbio3.1]{\begin{centering}
\includegraphics[scale=0.42]{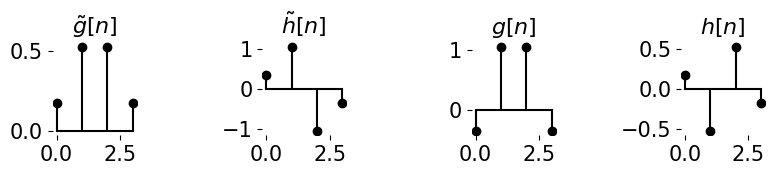}
\par\end{centering}
}
\par\end{centering}
\caption{(Left column) Wavelets, scaling functions and their corresponding impulse responses for bior3.1 (top row) and rbio3.1 (bottom row)  analysis and synthesis banks}

\label{fig:wavelets}
\end{figure}

A discrete wavelet system contains a pair of quadrature mirror filters
with a lowpass scaling function and a highpass mother wavelet. A one-dimensional ($\text{D}=1$) discrete
wavelet system with a continuous mother wavelet $\psi\in L^{2}(\mathbb{R})$ is realized by the impulse
responses of the scaling function and the wavelet as Finite Impulse
Response (FIR) filters $g\big[n\big]$ and $h\big[n\big]$ \cite{gadre2017multiresolution}. Figure
\ref{fig:wavelets} shows wavelets and scaling functions of the analysis
and synthesis bank of bior3.1 and rbio3.1 and their corresponding impulse
responses. These FIR filters are the building blocks of a two-band
perfect reconstruction filter bank for realizing fast discrete wavelet
systems. Figure \ref{fig:2BPRFB} shows a two-band perfect reconstruction
filter bank that operates on one-dimensional inputs, i.e., sequences
in $l^{2}(\mathbb{\mathbb{Z}})$. The analysis and synthesis bank
in orthogonal wavelet systems have identical lowpass and highpass
FIR filters but differ in biorthogonal wavelet systems. In biorthogonal
wavelet filter banks, $\tilde{g}\big[n\big]$ and $\tilde{h}\big[n\big]$
are the lowpass and highpass filters of the synthesis bank. The only
difference in the biorthogonal families like bior and rbio is the
interchange of the analysis and synthesis scaling and wavelets functions,
for example, in bior3.1 and rbio3.1.

\subsection{Circular convolution operators}

\begin{figure}[h]
\begin{centering}
\includegraphics[scale=0.6]{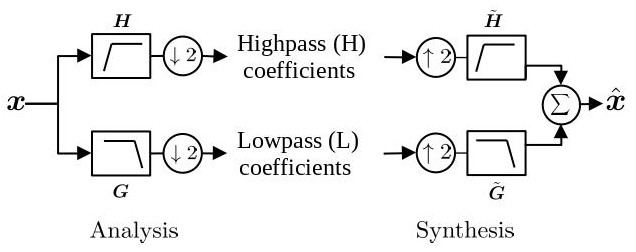}
\par\end{centering}
\caption{Two band perfect reconstruction filter bank}

\label{fig:2BPRFB}
\end{figure}

The four matrices in the two band perfect reconstruction filter bank
in Figure \ref{fig:2BPRFB} are --- (i) $\boldsymbol{G}$ is lowpass
analysis matrix, (ii) $\boldsymbol{H}$ is highpass analysis matrix,
(iii) $\tilde{\boldsymbol{G}}$ is lowpass synthesis matrix and (iv)
$\tilde{\boldsymbol{H}}$ is highpass synthesis matrix. These matrices
are operators for circular convolution, constructed by circular shifts
of the corresponding FIR filters $g\big[n-k\big]$, $h\big[n-k\big]$,
$\tilde{g}\big[n-k\big]$ and $\tilde{h}\big[n-k\big]$, where $g=\tilde{g}$
and $h=\tilde{h}$ for orthogonal wavelets.

\smallskip{}

\subsubsection{Analysis matrix}

For a lowpass analysis FIR filter of length $L$ and a input sequence
of length $N$, the circular convolution operator is a matrix $\boldsymbol{G}$
of shape $N\times N$. A downsampling by two $f_{(\downarrow2)}$
of the output the convolution is equivalent to getting rid of the
even rows of $\boldsymbol{G}$ to give $\frac{N}{2}\times{N}$
operator $f_{(\downarrow2)}\boldsymbol{G}$. Similarly, for the highpass
analysis FIR filter of same wavelet with convolution and downsampling
is a $\frac{N}{2}\times{N}$ operator $f_{(\downarrow2)}\boldsymbol{H}$.
The analysis matrix is given by equation (\ref{eq:AnalysisMatrix}), 

\begin{equation}
\boldsymbol{A}=\left[\begin{array}{c}
f_{(\downarrow2)}\boldsymbol{G}\\
f_{(\downarrow2)}\boldsymbol{H}
\end{array}\right]_{N\times N}\label{eq:AnalysisMatrix}
\end{equation}
where $\boldsymbol{A}$ is a $N\times N$ decimated circular convolution
operator formed by combining the lowpass and highpass decimated operators.

\smallskip{}

\subsubsection{Synthesis matrix}

The synthesis matrix $\boldsymbol{S}$ is another $N\times N$ decimated
circular convolution operator as given by equation (\ref{eq:SynthesisMatrix}),
\begin{equation}
\boldsymbol{S}=\left[\begin{array}{c}
f_{(\downarrow2)}\tilde{\boldsymbol{G}}\\
f_{(\downarrow2)}\tilde{\boldsymbol{H}}
\end{array}\right]_{N\times N}^{T}\label{eq:SynthesisMatrix}
\end{equation}
where $\tilde{\boldsymbol{G}}$ and $\tilde{\boldsymbol{H}}$ are
matrices formed by the lowpass and highpass synthesis FIR filters.
Equation (\ref{eq:SynthesisMatrix}) is a general representation for
both orthogonal and biorthogonal wavelets families where for orthogonal
wavelets, $\tilde{\boldsymbol{G}}=\boldsymbol{G}$, $\tilde{\boldsymbol{H}}=\boldsymbol{H}$
and thus $\boldsymbol{S}=\boldsymbol{A}^{T}$. 

\smallskip{}

\smallskip{}

A two-band perfect reconstruction discrete wavelet system for one-dimensional
inputs is given by the analysis equation (\ref{eq:DWToneSequence})
and the synthesis equation (\ref{eq:IDWToneSequence}),

\begin{align}
\boldsymbol{q} & =\text{{DWT}}\big(\boldsymbol{x}\big)\text{{ or, }}\boldsymbol{q}=\big(\boldsymbol{A}\boldsymbol{x}^{T}\big)^{T}\label{eq:DWToneSequence}\\
\nonumber \\\boldsymbol{x} & =\text{{IDWT}}\big(\boldsymbol{q}\big)\text{{ or, }}\boldsymbol{x}=\big(\boldsymbol{S}\boldsymbol{q}^{T}\big)^{T}\label{eq:IDWToneSequence}
\end{align}
where $\boldsymbol{A}$ and $\boldsymbol{S}$ are analysis and synthesis
matrices as defined in equations (\ref{eq:AnalysisMatrix}) and (\ref{eq:SynthesisMatrix}),
$\boldsymbol{x}$ is a input sequence and $\boldsymbol{q}$ has a
distinct lowpass and a highpass subband.

\vspace{1em}

\begin{example}
Given, a sequence $\boldsymbol{x}\in\mathbb{R}^{8}$ or $N=8$ and
FIR filters length $L=6$.
\end{example}
\begin{align*}
\text{\text{{LPF \& downsampling }}}f_{(\downarrow2)}\boldsymbol{G} & ={\left[\begin{array}{cccccccc}
g_{1} & g_{0} & 0 & 0 & g_{5} & g_{4} & g_{3} & g_{2}\\
g_{3} & g_{2} & g_{1} & g_{0} & 0 & 0 & g_{5} & g_{4}\\
g_{5} & g_{4} & g_{3} & g_{2} & g_{1} & g_{0} & 0 & 0\\
0 & 0 & g_{5} & g_{4} & g_{3} & g_{2} & g_{1} & g_{0}
\end{array}\right]_{\frac{N}{2}\times{N}}}\\
\\\text{{HPF \& downsampling }}f_{(\downarrow2)}\boldsymbol{H} & ={\left[\begin{array}{cccccccc}
h_{1} & h_{0} & 0 & 0 & h_{5} & h_{4} & h_{3} & h_{2}\\
h_{3} & h_{2} & h_{1} & h_{0} & 0 & 0 & h_{5} & h_{4}\\
h_{5} & h_{4} & h_{3} & h_{2} & h_{1} & h_{0} & 0 & 0\\
0 & 0 & h_{5} & h_{4} & h_{3} & h_{2} & h_{1} & h_{0}
\end{array}\right]_{\frac{N}{2}\times{N}}}
\end{align*}

\begin{align*}
\text{{Analysis matrix is }}\boldsymbol{A} & ={\left[\begin{array}{cccccccc}
g_{1} & g_{0} & 0 & 0 & g_{5} & g_{4} & g_{3} & g_{2}\\
g_{3} & g_{2} & g_{1} & g_{0} & 0 & 0 & g_{5} & g_{4}\\
g_{5} & g_{4} & g_{3} & g_{2} & g_{1} & g_{0} & 0 & 0\\
0 & 0 & g_{5} & g_{4} & g_{3} & g_{2} & g_{1} & g_{0}\\
h_{1} & h_{0} & 0 & 0 & h_{5} & h_{4} & h_{3} & h_{2}\\
h_{3} & h_{2} & h_{1} & h_{0} & 0 & 0 & h_{5} & h_{4}\\
h_{5} & h_{4} & h_{3} & h_{2} & h_{1} & h_{0} & 0 & 0\\
0 & 0 & h_{5} & h_{4} & h_{3} & h_{2} & h_{1} & h_{0}
\end{array}\right]_{{N}\times{N}}}\\
\\\text{{Similarly,}}\\
\text{{Synthesis matrix is }}\boldsymbol{S} & ={\left[\begin{array}{cccccccc}
\tilde{g}_{1} & \tilde{g}_{0} & 0 & 0 & \tilde{g}_{5} & \tilde{g}_{4} & \tilde{g}_{3} & \tilde{g}_{2}\\
\tilde{g}_{3} & \tilde{g}_{2} & \tilde{g}_{1} & \tilde{g}_{0} & 0 & 0 & \tilde{g}_{5} & \tilde{g}_{4}\\
\tilde{g}_{5} & \tilde{g}_{4} & \tilde{g}_{3} & \tilde{g}_{2} & \tilde{g}_{1} & \tilde{g}_{0} & 0 & 0\\
0 & 0 & \tilde{g}_{5} & \tilde{g}_{4} & \tilde{g}_{3} & \tilde{g}_{2} & \tilde{g}_{1} & \tilde{g}_{0}\\
\tilde{h}_{1} & \tilde{h}_{0} & 0 & 0 & \tilde{h}_{5} & \tilde{h}_{4} & \tilde{h}_{3} & \tilde{h}_{2}\\
\tilde{h}_{3} & \tilde{h}_{2} & \tilde{h}_{1} & \tilde{h}_{0} & 0 & 0 & \tilde{h}_{5} & \tilde{h}_{4}\\
\tilde{h}_{5} & \tilde{h}_{4} & \tilde{h}_{3} & \tilde{h}_{2} & \tilde{h}_{1} & \tilde{h}_{0} & 0 & 0\\
0 & 0 & \tilde{h}_{5} & \tilde{h}_{4} & \tilde{h}_{3} & \tilde{h}_{2} & \tilde{h}_{1} & \tilde{h}_{0}
\end{array}\right]_{{N}\times{N}}}
\end{align*}

\smallskip{}
\begin{align*}
\text{{The DWT of \ensuremath{\boldsymbol{x}} produces subbbands }}\ensuremath{\boldsymbol{q}} & =\text{{DWT}}\big(\boldsymbol{x}\big)\text{{ or, }}\boldsymbol{q=A}\boldsymbol{x}\\
\\\text{{Perfect reconstruction } \ensuremath{\boldsymbol{x}}} & =\text{{IDWT}}\big(\boldsymbol{q}\big)\text{{ or, }}\boldsymbol{x}=\boldsymbol{S}\boldsymbol{q}
\end{align*}

\smallskip{}

\subsection{DWT1D layer}

A DWT1D layer operates on input tensors of shape $(\text{{batch, length, channels}})$
and produces an output of shape $(\text{batch},\text{length}/2,2\times\text{channels})$
as described in Algorithm (\ref{algo:DWT1D}).
\begin{center}
\begin{algorithm}[H]
\caption{DWT1D layer}

\begin{enumerate}
\item Input $\boldsymbol{X}$ of shape $(\text{{batch, length, channels}})$.\\
\item Generate analysis matrix $\boldsymbol{A}$ using $\text{{length}}$
of input.\\
\item For each batched channel $\boldsymbol{x}\in\boldsymbol{X}$ of shape
$(\text{{batch, length}})$: $\boldsymbol{q}_{c}=\boldsymbol{A}\boldsymbol{x}^{T}$
\\
\item Stacking for all $c$ channels: $\boldsymbol{Q}:=\big(\boldsymbol{q}_{c}\big)_{\forall c}$
to a shape $(\text{{batch, length, channels}})$.\\
\item Group subbands and return an output $\boldsymbol{Q}^{(\text{{grouped}})}$
of shape $(\text{batch},\text{length}/2,,2\times\text{channels})$\\
\begin{align*}
\text{{mid}} & =\text{{int}}(Q.\text{{shape}}[1]/2\\
\text{{L}} & =Q[:,:\text{{mid}},:]\\
\text{{H}} & =Q[:,\text{{mid}}:,:]\\
\boldsymbol{Q}^{(\text{{grouped}})} & =\text{{Concatenate}}([\text{{L}},\text{{H}}],\text{{axis}}=-1)
\end{align*}
\end{enumerate}
\label{algo:DWT1D}
\end{algorithm}
\par\end{center}

\smallskip{}

\subsection{IDWT1D layer}

An IDWT1D layer operates on input tensors of shape $(\text{batch},\text{length}/2,2\times\text{channels})$
and produces an output of shape $(\text{{batch, length, channels}})$
as described in Algorithm (\ref{algo:IDWT1D}).
\begin{center}
\begin{algorithm}[H]
\caption{IDWT1D layer}

\begin{enumerate}
\item Input $\boldsymbol{Q}^{(\text{{grouped}})}$ of shape $(\text{batch},\text{length}/2,2\times\text{channels})$\\
\item Ungroup the subbands to get $\boldsymbol{Q}$ of shape $(\text{{batch, length, channels}})$\\
\item Generate synthesis matrix $\boldsymbol{S}$ using $\text{{length}}$
of $\boldsymbol{Q}$.\\
\item For each batched channel $\boldsymbol{q}\in\boldsymbol{Q}$ of shape
$(\text{{batch, length}})$: $\boldsymbol{x}=\boldsymbol{S}\boldsymbol{q}^{T}$,
i.e., a perfect reconstruction, where, $\boldsymbol{S}=\boldsymbol{A}^{T}$
for orthogonal wavelets\\
\item Layer output (perfect reconstruction): $\boldsymbol{X}:=\big(\boldsymbol{x}_{c}\big)_{\forall c}$
is of shape $(\text{{batch, length, channels}})$
\end{enumerate}
\label{algo:IDWT1D}
\end{algorithm}
\par\end{center}

\smallskip{}

\section{Higher dimensional discrete wavelet systems}

In sequences (1D), the DWT1D applies to the only independent variable.
To achieve high-dimensional DWT, the DWT1D needs to be applied separably
to all the independent variables one after the other. For example,
the DWT of an image (2D) is a row-wise DWT1D followed by a column-wise
DWT1D. Similarly, the reconstruction is column-wise IDWT1D followed
by a row-wise IDWT1D.

\smallskip{}

\subsection{Two-dimensional discrete wavelet system}

\begin{figure}[H]
\begin{centering}
\includegraphics[scale=0.55]{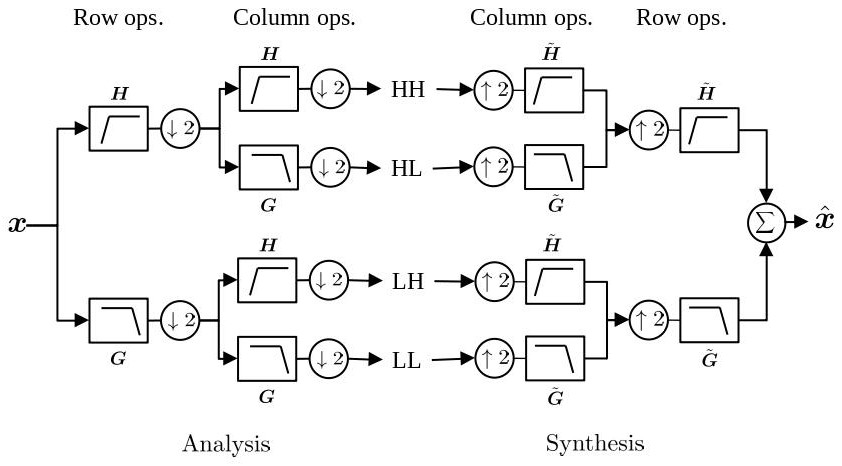}
\par\end{centering}
\caption{Separable DWT2D perfect reconstruction filter bank}

\label{fig:2BPRFB-2}
\end{figure}

\begin{figure}[h]
\begin{centering}
\includegraphics[scale=0.38]{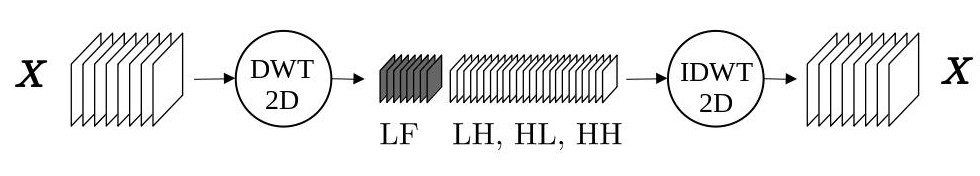}
\par\end{centering}
\caption{DWT2D decomposition layer and perfect reconstruction with IDWT2D layer of a tensor $\boldsymbol{X}$}

\label{fig:2DDWTlayer}
\end{figure}

The pixel values in an image are a function of two independent spatial
axes. A DWT2D filter bank is a separable transform with row-wise
filtering followed by column-wise filtering that yields four subbands
- LL, LH, HL and HH. A two-dimensional discrete wavelet system is
given by,

\begin{align}
\boldsymbol{q} & =\text{{DWT}}\big(\boldsymbol{x}\big):=\boldsymbol{A}(\boldsymbol{A}\boldsymbol{x}_{021})_{021}^{T}\label{eq:Analysis2D}\\
\nonumber \\\boldsymbol{x} & =\text{{IDWT}}\big(\boldsymbol{q}\big):=\boldsymbol{x}=\boldsymbol{S}(\boldsymbol{S}\boldsymbol{q}_{021}^{T})_{021}^{T}\label{eq:Synthesis2D}
\end{align}
\begin{figure}[H]
\begin{centering}
\includegraphics[scale=0.45]{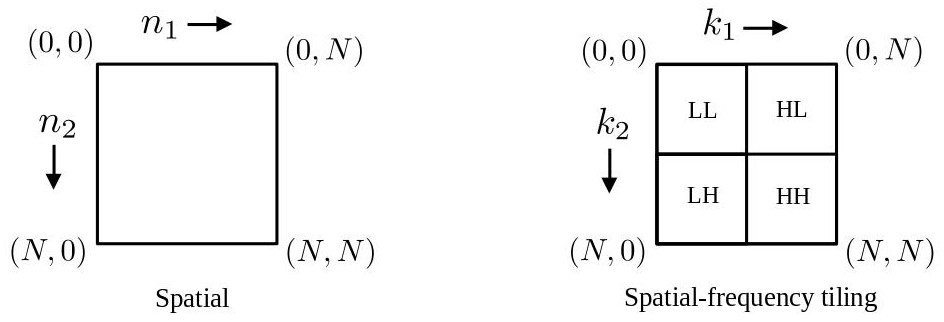}
\par\end{centering}
\caption{An image of shape $N\times N$ (left) and spatial-frequency tiled
subbands (right) after a level-1 DWT2D decomposition}

\label{fig:subdband2d}
\end{figure} 
where, the indices $021$ in $\boldsymbol{x}_{021}$ denote a transpose
of axis $[0,1,2]$ to $[0,2,1]$, with $0$ as batch index, $1$ and
$2$ are height and width. The matrices $\boldsymbol{A}$ and $\boldsymbol{S}$ are the same analysis and synthesis matrices as defined in equations (\ref{eq:AnalysisMatrix}) and (\ref{eq:SynthesisMatrix}). Figure (\ref{fig:subdband2d}) shows an $N\times N$ image and its four localized spatial-frequency subbands after DWT. Here, the low-frequency band is LL, and the other three are high-frequency subbands representing horizontal, vertical and diagonal features. Figure (\ref{fig:2BPRFB-2}) illustrates a separable 2D DWT perfect reconstruction filter bank.
 The 2D layers operate on a batch of multichannel tensors of shape $(\text{{batch, height, width, channels}})$,
where each image is of shape height and width. Figure (\ref{fig:2DDWTlayer})
illustrates the input, output and perfect reconstruction by DWT2D and IDWT2D layers. The Multiresolution Encoder-Decoder Convolutional
Neural Network in \cite{tarafdar2025multiresolution} uses these forward
and inverse layers for semantic segmentation. 

\cleardoublepage{}
\subsubsection{DWT2D layer}

A DWT2D layer operates on input tensors of shape $(\text{{batch, height, width, channels}})$
and produces an output of shape $(\text{{batch, height}}/2,\text{{width}}/2,4\ensuremath{\times}\text{{channels}})$
as described in Algorithm \ref{algo:DWT2D}. 
\begin{center}
\begin{algorithm}[H]
\caption{DWT2D layer}

\begin{enumerate}
\item Input $\boldsymbol{X}$ of shape $(\text{{batch, height, width, channels}})$.\\
\item Generate analysis matrix $\boldsymbol{A}$ using $\text{{height}}$
and $\text{{width}}$ of input.\\
\item For each batched channel $\boldsymbol{x}_{c}\in\boldsymbol{X}$ of
shape $(\text{{batch, height, width}})$:\\
(omitting suffix $c$ in $\boldsymbol{x}$ below for simplicity of
notation)
\begin{enumerate}
\item Batched row wise DWT,\\
\\
$\boldsymbol{A}\boldsymbol{x}_{021}^{T}:=\text{{Einsum}\ensuremath{\big(}}ij,bjk\rightarrow bik\big)$\\
\item Batched column wise DWT,\\
\\
$\boldsymbol{A}(\boldsymbol{A}\boldsymbol{x}_{021}^{T})_{021}^{T}:=\text{{Einsum}\ensuremath{\big(}}ij,bjk\rightarrow bik\big)$
\\
\\
or, equivalently, DWT of a batched channel $\boldsymbol{x}$ is, \\
\\
$\boldsymbol{q}_{c}=\text{{DWT}}\big(\boldsymbol{x}\big)\text{{ or, }}\boldsymbol{q}_{c}=\boldsymbol{A}(\boldsymbol{A}\boldsymbol{x}_{021})_{021}^{T}$\\
\end{enumerate}
\item Stacking for all $c$ channels, $\boldsymbol{Q}:=\big(\boldsymbol{q}_{c}\big)_{\forall c}$
to a shape $(\text{{batch, height, width, channels}})$.\\
\item Group subbands and return an output $\boldsymbol{Q}^{(\text{{grouped}})}$
of shape $(\text{{batch, height/2, width/2, 4\ensuremath{\times}channels}})$\\
\begin{align*}
\text{{mid}} & =\text{{int}}(Q.\text{{shape}}[1]/2)\\
\text{{LL}}= & Q[:,:\text{{mid}},:\text{{mid}},:]\\
\text{{LH}} & =Q[:,\text{{mid}}:,:\text{{mid}},:]\\
\text{{HL}} & =Q[:,:\text{{mid}},\text{{mid}}:,:]\\
\text{{HH}} & =Q[:,\text{{mid}}:,\text{{mid}}:,:]\\
\boldsymbol{Q}^{(\text{{grouped}})} & =\text{{Concatenate}}([\text{{LL}},\text{{LH}},\text{{HL}},\text{{HH}}],\text{{axis}}=-1)
\end{align*}
\end{enumerate}
\label{algo:DWT2D}
\end{algorithm}
\smallskip{}
\par\end{center}

\par \textbf{Einstein summations}: 
Here, the matrix multiplications in each separable transform are performed optimally with Einstein summations or $\text{Einsum}$. $\text{Einsum}$ provides a powerful and concise way to perform tensor operations such as matrix multiplication, dot products, element-wise operations, and more. It simplifies complex calculations by eliminating the need for explicit loops, using repeated indices to imply summation. 

\cleardoublepage{}
\subsubsection{IDWT2D layer}

An IDWT2D layer operates on input tensors of shape $(\text{batch}$, $\text{height}/2$, $\text{width}/2$, $4\ensuremath{\times}\text{channels})$
and produces an output of shape $(\text{{batch, height, width, channels}})$
as described in Algorithm \ref{Algo:IDWT2DLayer}. 
\begin{algorithm}[H]
\caption{IDWT2D layer}

\begin{enumerate}
\item Input $\boldsymbol{Q}^{(\text{{grouped}})}$ of shape $(\text{{batch, height/2, width/2, 4\ensuremath{\times}channels}})$\\
\item Ungroup the subbands to get $\boldsymbol{Q}$ of shape $(\text{{batch, height, width, channels}})$\\
\item Generate synthesis matrix $\boldsymbol{S}$ using $\text{{height}}$
and $\text{{width}}$ of $\boldsymbol{Q}$.\\
\item For each batched channel $\boldsymbol{q}_{c}\in\boldsymbol{Q}$ of
shape $(\text{{batch, height, width}})$:\\
(omitting suffix $c$ in $\boldsymbol{q}$ below for simplicity of
notation)
\begin{enumerate}
\item Batched row wise IDWT ,\\
\\
$\boldsymbol{S}\boldsymbol{q}_{021}^{T}:=\text{{Einsum}\ensuremath{\big(}}ij,bjk\rightarrow bik\big)$\\
\item Batched column wise IDWT,\\
\\
$\boldsymbol{S}(\boldsymbol{S}\boldsymbol{q}_{021}^{T})_{021}^{T}:=\text{{Einsum}\ensuremath{\big(}}ij,bjk\rightarrow bik\big)$\\
\\
or equivalently, a perfect reconstruction, \\
\\
$\boldsymbol{x}=\text{{IDWT}}\big(\boldsymbol{q}\big)\text{{ or, }}\boldsymbol{x}=\boldsymbol{S}(\boldsymbol{S}\boldsymbol{q}_{021}^{T})_{021}^{T}$\\
\\
where, $\boldsymbol{S}=\boldsymbol{A}^{T}$ for orthogonal wavelets 
\end{enumerate}
\item Layer output (perfect reconstruction) $\boldsymbol{X}:=\big(\boldsymbol{x}_{c}\big)_{\forall c}$
is of shape $(\text{{batch, height, width, channels}})$
\end{enumerate}
\label{Algo:IDWT2DLayer}
\end{algorithm}
\smallskip{}

\subsection{Three-dimensional discrete wavelet system}

A three-dimensional (3D) discrete wavelet system for a 3D input $\boldsymbol{x}$
is given by,

\smallskip{}
\begin{align}
\boldsymbol{q} & =\text{{DWT}}\big(\boldsymbol{x}\big):=\left[\boldsymbol{A}(\boldsymbol{A}(\boldsymbol{A}\boldsymbol{x}_{0213}^{T})_{0213}^{T})_{0132}^{T}\right]{}_{0132}^{T}\label{eq:Analysis3D}\\
\nonumber \\\boldsymbol{x} & =\text{{IDWT}}\big(\boldsymbol{q}\big):=\left[\boldsymbol{S}(\boldsymbol{S}(\boldsymbol{S}\boldsymbol{q}_{0132}^{T})_{0132}^{T})_{0213}^{T}\right]_{0213}^{T}\label{eq:Synthesis3D}
\end{align}
\smallskip{}
where, $\boldsymbol{x}_{0213}^{T}$ denotes transpose of axis $[0,1,2,3]$
to $0,2,1,3]$ with 0 as batch index; 1,2 and 3 are height, width
and depth. The matrices $\boldsymbol{A}$ and $\boldsymbol{S}$ are
the same analysis and synthesis matrices as defined in equations (\ref{eq:AnalysisMatrix})
and (\ref{eq:SynthesisMatrix}). The DWT3D and IDWT3D layers operate
on batched, multichannel tensors of shape $(\text{{batch, height, width, depth, channels}})$,
where each cube is transformed separably by equations (\ref{eq:Analysis3D})
and (\ref{eq:Synthesis3D}).

\cleardoublepage{}

\subsubsection{DWT3D layer}

A DWT3D layer operates on input tensors of shape ($\text{batch}$, $\text{height}$, $\text{width}$, $\text{depth}$, $\text{channels})$
and produces an output of shape $(\text{batch},\text{height}/2,\text{width}/2,\text{depth}/2,8\times\text{channels})$
as described in Algorithm \ref{Algo:DWT3Dlayer}.

\begin{algorithm}[H]
\caption{DWT3D layer}

\begin{enumerate}
\item Input $\boldsymbol{X}$ of shape $(\text{{batch, height, width, depth, channels}})$.
\item Generate analysis matrix $\boldsymbol{A}$ using $\text{{height}}$,
$\text{{width}}$ and $\text{{depth}}$of input.
\item For each batched channel $\boldsymbol{x}_{c}\in\boldsymbol{X}$ of
shape $(\text{{batch, height, width, depth}})$:\\
(omitting suffix $c$ in $\boldsymbol{x}$ below for simplicity of
notation)
\begin{enumerate}
\item Row-wise operations,\\
\\
$\boldsymbol{A}\boldsymbol{x}_{0213}^{T}:=\text{{Einsum}}\big(ij,bjkl\rightarrow bikl\big)$\\
\item Column-wise operations, \\
\\
$\boldsymbol{A}\left(\boldsymbol{A}\boldsymbol{x}_{0213}^{T}\right)_{0213}^{T}:=\text{{Einsum}}\big(ij,bjkl\rightarrow bikl\big)$\\
\item Depth-wise operations, \\
\\
$\boldsymbol{A}(\boldsymbol{A}(\boldsymbol{A}\boldsymbol{x}_{0213}^{T})_{0213}^{T})_{0132}^{T}:=\text{{Einsum}}\big(ik,bjkl\rightarrow bjil\big)$\\
\\
Therefore the forward transform of $\boldsymbol{x}$ yield DWT analysis
coefficients\\
\\
$\boldsymbol{q}_{c}:=\text{{DWT}}\left(\boldsymbol{x}\right)=\left[\boldsymbol{A}(\boldsymbol{A}(\boldsymbol{A}\boldsymbol{x}_{0213}^{T})_{0213}^{T})_{0132}^{T}\right]{}_{0132}^{T}$  \\
\end{enumerate} 
\item Stacking for all $c$ channels, $\boldsymbol{Q}:=\big(\boldsymbol{q}_{c}\big)_{\forall c}$
to a shape $(\text{{batch, height, width, depth, channels}})$.\\
\item Group subbands and return an output $\boldsymbol{Q}^{(\text{{grouped}})}$
of shape $(\text{{batch, height/2, width/2, depth/2, 8\ensuremath{\times}channels}})$,
\begin{align*}
\text{mid} & =\text{int}(Q.\text{shape}[2]/2)\\
\text{LLL} & =Q[:,:\text{mid},:\text{mid},:\text{mid},:]\\
\text{LLH} & =Q[:,\text{mid}:,:\text{mid},:\text{mid},:]\\
\text{LHL} & =Q[:,:\text{mid},\text{mid}:,:\text{mid},:]\\
\text{LHH} & =Q[:,\text{mid}:,\text{mid}:,:\text{mid},:]\\
\text{HLL} & =Q[:,:\text{mid},:\text{mid},\text{mid}:,:]\\
\text{HLH} & =Q[:,\text{mid}:,\text{mid}:,\text{mid}:,:]\\
\text{HHL} & =Q[:,:\text{mid},\text{mid}:,\text{mid}:,:]\\
\text{HHH} & =Q[:,\text{mid}:,\text{mid}:,\text{mid}:,:]\\
\boldsymbol{Q}^{(\text{{grouped}})} & =\text{Concatenate([\text{LLL}, \text{LLH}, \text{LHL}, \text{LHH}, \text{HLL}, \text{HLH}, \text{HHL}, \text{HHH]}, \text{axis}=-1)}
\end{align*}
\end{enumerate}
\label{Algo:DWT3Dlayer}
\end{algorithm}

\cleardoublepage{}

\subsubsection{IDWT3D layer}

An IDWT3D layer operates on input tensors of shape ($\text{batch}$, $\text{height}/2$,$\text{width}/2$, $\text{depth}/2$,$8\times\text{channels}$)
and produces an output of shape $(\text{{batch, height, width, depth, channels}})$
as described in Algorithm \ref{Algo:IDWT3Dlayer}.

\begin{algorithm}[H]
\caption{IDWT3D layer}

\begin{enumerate}
\item Input $\boldsymbol{Q}^{(\text{{grouped}})}$ of shape $(\text{batch},\text{height}/2,\text{width}/2,\text{depth}/2,8\times\text{channels})$\\
\item Ungroup to get $\boldsymbol{Q}$ of shape $(\text{{batch, height, width, depth, channels}})$\\
\item Generate synthesis matrix $\boldsymbol{S}$ using $\text{{height}}$,
$\text{{width}}$ and $\text{{depth}}$ of $\boldsymbol{Q}$.\\
\item For each batched channel $\boldsymbol{q}_{c}\in\boldsymbol{Q}$ of
shape $(\text{{batch, height, width, depth}})$:\\
(omitting suffix $c$ in $\boldsymbol{q}$ below for simplicity of
notation)
\begin{enumerate}
\item Row-wise operations,\\
\\
$\boldsymbol{S}\boldsymbol{q}_{0132}^{T}:=\text{{Einsum}}\big(ik,bjkl\rightarrow bjil\big)$\\
\item Column-wise operations,\\
\\
$\boldsymbol{S}(\boldsymbol{S}\boldsymbol{q}_{0132}^{T})_{0132}:=\text{{Einsum}}\big(ij,bjkl\rightarrow bikl\big)$\\
\item Depth-wise operations,\\
\\
$\boldsymbol{S}(\boldsymbol{S}(\boldsymbol{S}\boldsymbol{q}_{0132}^{T})_{0132}^{T})_{0213}^{T}:=\text{{Einsum}}\big(ij,bjkl\rightarrow bikl\big)$\\
\\
or equivalently, a perfect reconstruction,\\
\\
$\boldsymbol{x}_{c}:=\text{{IDWT}}\left(\boldsymbol{q}\right)=\left[\boldsymbol{S}(\boldsymbol{S}(\boldsymbol{S}\boldsymbol{q}_{0132}^{T})_{0132}^{T})_{0213}^{T}\right]_{0213}^{T}$\\
\\
where, $\boldsymbol{S}=\boldsymbol{A}^{T}$ for orthogonal wavelets.
\end{enumerate}
\item Layer output $\boldsymbol{X}:=\big(\boldsymbol{x}_{c}\big)_{\forall c}$
is of shape $(\text{{batch, height, width, depth, channels}})$
\\
(Perfect reconstruction) 
\end{enumerate}
\label{Algo:IDWT3Dlayer}
\end{algorithm}

\smallskip{}

\smallskip{}

\smallskip{}

In general, a seamless realization of fast $\text{D}$-dimensional
DWT and IDWT is possible by extending the above separable method to
all independent $N$ axes one after the other. The number of subbands
will be equal to $2^{\text{{D}}}$ for a $\text{{D}}$ dimensional
DWT. For example, sequences $(\text{{D}}=1)$ yield two subbands,
images $(\text{{D}}=2)$ yield four subbands, three-dimensional inputs
$(\text{{D}}=3)$ with voxels yield eight subbands, etc.\smallskip{}

\section{Multilevel wavelet filter banks}

\begin{figure}[H]
\begin{centering}
\includegraphics[scale=0.32]{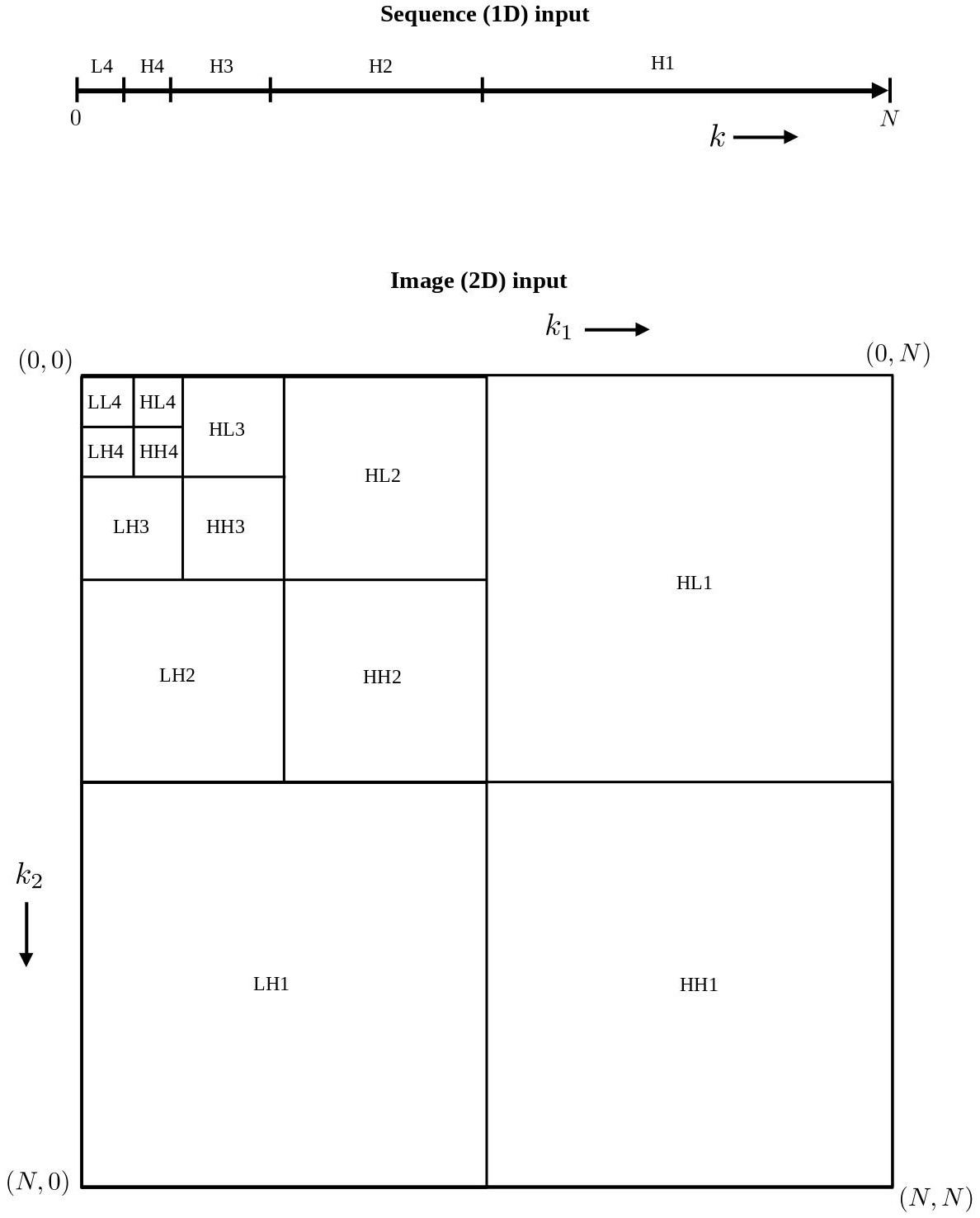}
\par\end{centering}
\caption{Natural-frequency tiling by DWT level 4 decomposition of a sequence
of length $N$ (top) and image of shape $N\times N$ (bottom)}

\label{fig:MultilevelDWTtiling}
\end{figure}

The discussed DWT and IDWT layers are building blocks in the construction of multilevel DWT filter banks. Figure \ref{fig:MultilevelDWTtiling} shows the partitioning of the 1D frequency axis and the tiling of the 2D frequency plane using a level-4 1D and 2D DWT. The multilevel DWT successively decomposes the low-frequency feature. If the high-frequency features are also decomposed successively, then we get a Wavelet Packet Transform (WPT) filter bank.

% \bibliographystyle{ieeetr}
% \bibliography{references}

\end{document}